\documentclass[journal,12pt,onecolumn,draftcls]{IEEEtran}
\usepackage{float}
\usepackage{comment}
\usepackage{lscape}
\usepackage[table]{xcolor}
\usepackage{cite}      
\usepackage{graphicx}  
\usepackage{psfrag}    
\usepackage{placeins}
\usepackage{lineno}
\usepackage{url}
\usepackage{subcaption}
\usepackage{times}
\usepackage{textcomp}
\usepackage{amsmath} 
\usepackage{amsfonts,amsthm,amssymb,mathtools}
\usepackage{balance}
\usepackage[normalem]{ulem}
\usepackage{xcolor}

\usepackage{cuted}
\usepackage{amsmath,bm}
\usepackage{cleveref}

\DeclareMathOperator{\tpi}{\tau_{b}}
\DeclareMathOperator{\tpo}{\tau_{b_i}}
\DeclareMathOperator{\tpone}{\tau_{b_1}}
\DeclareMathOperator{\tp2}{\tau_{b_2}}

\DeclareMathOperator{\C}{\mathcal{C}}
\DeclareMathOperator{\R}{\mathcal{R}}

\DeclareMathOperator{\xw}{\mathbf{x}}
\DeclareMathOperator{\xc}{\mathbf{x}^{\C}}
\DeclareMathOperator{\xch}{\mathbf{\bar{x}}^{\C}}
\DeclareMathOperator{\xwh}{\mathbf{\bar{x}}}
\DeclareMathOperator{\xe}{x_k}
\DeclareMathOperator{\ye}{y_k}
\DeclareMathOperator{\vx}{v_{x_k}}
\DeclareMathOperator{\vy}{v_{y_k}}
\DeclareMathOperator{\ze}{z_k}

\usepackage{multicol}
\usepackage{xr-hyper}

\begin{document}
%\linenumbers
\title{Fusion of Inverse Synthetic Aperture Radar and Camera Images for Automotive Target Tracking}
\author{Shobha~Sundar~Ram,~\IEEEmembership{Senior Member,~IEEE}
\thanks{The authors are with Indraprastha Institute of Information Technology Delhi (email: shobha@iiitd.ac.in).}}
% make the title area
\maketitle

\begin{abstract}
Automotive targets undergoing turns in road junctions offer large synthetic apertures over short dwell times to automotive radars that can be exploited for obtaining fine cross-range resolution. Likewise, the wide bandwidths of the automotive radar signal yield high-range resolution profiles. Together, they are exploited for generating inverse synthetic aperture radar (ISAR) images that offer rich information regarding the target vehicle's size, shape, and trajectory which is useful for object recognition and classification. However, a key requirement for ISAR is translation motion compensation and estimation of the turning velocity of the target. State-of-the-art algorithms for motion compensation trade-off between computational complexity and accuracy. An alternative low complexity method is to use an additional sensor for tracking the target motion. In this work, we propose to exploit  computer vision algorithms to identify the radar target object in the sensor field-of-view (FoV) with high accuracy. Further, we propose to track the target vehicle's motion through fusion of vision and radar data. Vision data facilitates the accurate estimation of the lateral position of the target and its lateral velocity which complements the radar’s capability of accurate estimation of range and radial velocity. Through simulations and experimental evaluations with a monocular camera and Texas Instrument’s millimeter wave radar, we demonstrate the effectiveness of sensor fusion for accurate target tracking for translational motion compensation and generation of high quality ISAR images.  
\end{abstract}
%\IEEEpeerreviewmaketitle
%\begin{comment}

\section{Introduction}
Millimeter wave automotive radars transmit wideband frequency modulated continuous waveforms that enable the generation of high range-resolution profiles \cite{hasch2012millimeter}. When combined with wide radar antenna apertures, high resolution two-dimensional (2D) images along the top-view (or bird's eye view) of an object can be realized at short ranges. Large radar apertures can be realized through the use of multiple antenna elements and corresponding receiver channels. These electrically large antenna apertures can be fabricated with a small physical footprint at millimeter wave frequencies. However, the cost and complexity of the multichannel system are significant. Further, fully or even semi-autonomous vehicles are envisaged to have multiple radars mounted all around the vehicle to provide $360^{\circ}$ coverage which further scales the cost. Inverse synthetic aperture radar (ISAR) imaging presents a low complexity alternative to large antenna arrays. When a target rotates or turns while remaining in the radar beam, the aspect presented by the target provides a large synthetic aperture which can be exploited for fine cross-range resolution determined by the total aspect and the target dwell time \cite{chen2014inverse}. Further, these turns are commonly encountered at road junctions.  

ISAR imaging has been researched for the detection and classification of several airborne and waterborne bodies \cite{martorella2009automatic,vespe2007automatic,park2011construction,wang2015inverse}. More recently SAR and ISAR imaging have been examined for ground-based vehicles \cite{Sun2007adaptive,kulpa2013experimental,li2015wide,essen2008high,gishkori2021imaging,pandey2020database,pandey2022classification}. These works have demonstrated that wideband synthetic aperture data result in images that provide rich information on the size, geometry, and trajectory undertaken by the target vehicle along the top-view that can be exploited for object recognition. Several methods have been explored for generating cross-range information along elevation. One possibility is to exploit the target rotation about the horizontal axis (roll or pitch) along with the yaw. This method, while suited for certain types of airborne targets, is not practical for automotive targets where the only components that undergo roll motion are the wheels. The second possibility is to perform beamforming along the elevation by incorporating multiple receiver elements along the elevation dimension. However, this again increases the complexity of the radar architecture. The simplest alternative, that is investigated in this work, is \emph{interferometric ISAR} where we estimate the height of target features using differential interferometry between the ISAR images generated at two receiver elements spaced along the height dimension. The results show that the target features are of very low (and similar) elevation angles even at short ranges of the vehicle from the radar. 

There are some key challenges associated with  ISAR imaging. First, the fundamental principle of ISAR imaging is that the Doppler frequency shift in the radar received signal should arise from the rotational motion of the target instead of the translational motion. Therefore, the translational motion of the target has to be accurately compensated to ensure that the images are focused. Second, the rotational velocity of the target has to be correctly estimated in order to map the Doppler frequency axis to the cross-range axis. Several algorithms in radar literature have been explored for \emph{blind} motion compensation when the target trajectory is not known to the radar. The traditional methods involved tracking the  dominant scatterers in the radar data either in the range domain \cite{wu1995translational,itoh1996motion} or along the Doppler domain \cite{wang1998isar} and performing range alignment or auto-focusing of the ISAR images. Later, Keystone transform was explored to perform higher order motion compensation for ISAR images \cite{xing2004migration,li2017robust}.
More recently, particle swarm optimization \cite{liu2015adaptive} and sparsity-based techniques \cite{shao2020high} have been explored to perform higher-order motion compensation. These algorithms have typically traded off between accuracy in motion compensation and computational cost. Alternatively, the use of secondary radars that provides ground truth target state information has also been suggested to reduce the computational complexity \cite{li2015wide}. Due to the ubiquity of cameras for assisted and autonomous driving applications, the use of vision data for enhancing radar imaging forms a natural solution for automotive applications that we explore in this work. 

Concurrent with the research of automotive radar imaging of road users, computer vision algorithms are being developed for road object recognition and classification \cite{viola2001rapid,dollar2009integral,dollar2014fast}. A monocular camera offers frontal images as opposed to the top-view images obtained from radar. These images are characterized by excellent lateral resolution and corresponding lateral velocity resolution but require more complex processing for recovering depth information \cite{saxena2005learning}. Even stereo cameras can provide acceptable estimates of the depth/range only up to a distance of a few meters \cite{scharstein2002taxonomy}. On the other hand, while the range and radial velocity estimation by radar is accurate, the lateral velocity estimation is very poor. Thus the radar and camera apertures yield information that presents complementary features for object tracking. In this work, we propose the fusion of monocular vision data with radar data to enable target state estimation - position, velocity and turn rate. Then, we exploit the estimated target trajectory information for performing motion compensation on radar data in order to realize high-resolution ISAR images. 

Several works in the last few years have explored the fusion of data from two or more sensors for supporting advanced driver assistance system functionalities. Sensor fusion offers several advantages \cite{chavez2015multiple}. First, are the advantages of redundancy in the event of any single sensor malfunction or breakdown. For example, camera images are affected by weather and light conditions. Radars can work 24×7, in all weather conditions, and to a limited extent in non-line-of-sight conditions as well due to the presence of multipath \cite{ram2010simulation,vishwakarma2020micro}. Secondly, multiple sensors are useful for reducing the errors associated with the estimation of target parameters. Radars and cameras, in particular, as mentioned earlier, offer complementary information that when fused together increase the statistical information of a target. Cameras offer vital information on the size, shape, color, and texture of various objects on the road and hence are extremely effective for object detection and classiﬁcation \cite{dollar2014fast}. Radar signatures, on the other hand, are not always intuitive and require a trained operator for interpretation. Further, the mitigation of clutter and multipath in radar images can be challenging \cite{vishwakarma2020mitigation,ram2021sparsity}. 

Many works over the last two decades have explored the fusion of radar and vision data. One of the earliest works explored the use of depth estimates provided by radar data to enhance the target segmentation in the camera image \cite{fang2002depth}. Later the authors in \cite{alessandretti2007vehicle} combined radar and vision data to detect guard rails in the ego vehicle's environment. In \cite{wu2009collision}, the fused vision and radar data were used to estimate the position, velocity, and orientation of an oncoming vehicle to enable collision avoidance. In all of these works, low-resolution radar detections - a single detection per target - are obtained for automotive targets which are subsequently used for localization. The key distinction of our work from the prior art in radar-vision fusion is that we are proposing a method to fuse \emph{high resolution} radar images with camera data. 

The paper is organized in the following manner. In the following section, we present the algorithms for processing radar data in order to obtain 3D ISAR images; algorithms for the camera data processing for estimating the position of the target vehicle; and the radar-camera fusion framework for estimating the target position and focusing the radar images. In Section III, we present the simulation setup for generating radar and camera data from a cuboid model of a target vehicle. The simulations model the sensors with a finite probability of detection, occlusion effects and shadowing, and false alarms from clutter. The simulation results demonstrate the effectiveness of focusing high-resolution radar images from monocular vision data through comparison with ground truth radar images. The algorithms are further validated with experimental data gathered from both the radar and the camera in Section IV. Finally, we conclude the paper with a discussion of the key insights and scope for future work in Section V.

\emph{Notation:} Lowercase Latin and Greek letters denote scalar variables while boldface lowercase and uppercase letters represent vectors and matrices respectively. The homogeneous representation of a 2D or 3D point is $\mathbf{\bar{x}}=[\mathbf{x};1]$.
\section{Theory}
\begin{figure*}
    \centering
    \includegraphics[width=6in,height=2in]{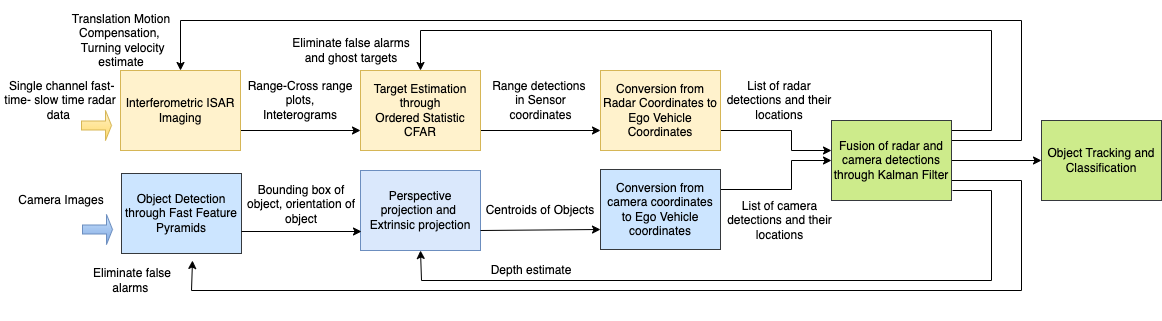}
    \caption{System framework of radar and camera data processing and fusion for automotive target imaging}
    \label{fig:JSTSP_SARsystem}
\end{figure*}

In self-driving cars, multiple millimeter wave radars and cameras are mounted all around the chassis of the ego vehicle for complete coverage of the road conditions. In this work, we consider a single forward facing low complexity three element radar with a single transmitter phase synchronized with two receivers; and a single monocular forward-facing camera in close proximity to the radar so that there is a large overlap in their fields-of-view (FoV). We consider two distinct coordinate frames corresponding to the radar/ego vehicle ($\R$) and the camera ($\C$) with a shared common ground plane at $z=0$. The misalignment between the X and Y axes between the two frames is mapped through the Euler rotation matrix, denoted by $\mathbf{R}^{\C\R}$ (from $\C$ to $\R$), and the translational matrix denoted by $\mathbf{t}^{\C\R}$. We consider a single automotive target scenario where the target is detected jointly by the radar and camera. The detections from the two sensors are time stamped to provide updates at a low frame rate (a few samples per second) indexed by $k$. The dual sensor data are processed and subsequently fused to estimate the target as described in the section below. Subsequently, the target yaw rate estimated by the sensor fusion is used to generate the ISAR images. The entire data processing chain for camera and radar data is presented in Fig.\ref{fig:JSTSP_SARsystem}. 
In this section, we first describe the radar processing steps followed by the camera processing steps and finally the fusion.
\subsection{Interferometric ISAR Imaging of Extended Targets}
The automotive radar transmitter and the two receivers are assumed to be configured in an L formation with the receivers displaced from each other by half-wavelength along the $z$ axis. The radar transmits linear frequency modulated waveforms modeled as \par\noindent\small
\begin{align}
\mathbf{s}_{tx}(\tau) = \text{rect}\left(\frac{\tau}{T_{PRI}}\right)e^{j2\pi f_c\tau}e^{j\pi\beta \tau^2},
\end{align}\normalsize
where $\tau$ denotes the fast time samples within a pulse repetition interval of $T_{PRI}$. The carrier frequency of the transmitted signal is $f_c$ and $\beta$ is the chirp rate of the signal. At short ranges ($\le$50m), the automotive radar target appears as an extended target with $b=1:B$ point scatterers distributed across several range and Doppler bins. The radar transmit signal is scattered by each $b^{th}$ point scatterer and received by the $i^{th}, i= 1,2$ radar receiver element after a two-way round-trip time delay, $\tau_{b_i}$, which we assume is within $T_{PRI}$ to avoid range ambiguity. 
\begin{figure}
    \centering
\includegraphics[width=3.5in,height=1.25in]{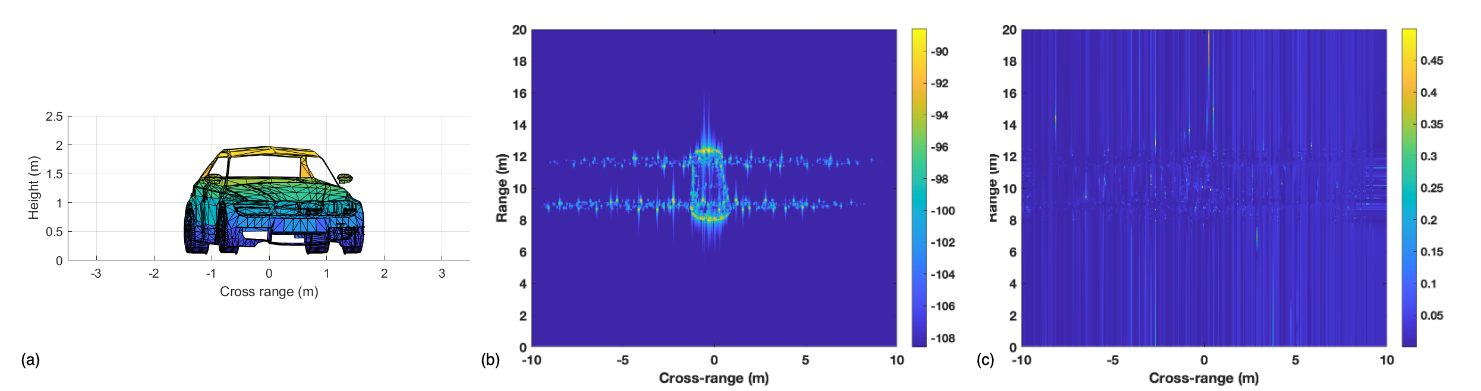}
    \caption{(a) Extended radar target model of a full-sized car; (b) ISAR image after perfect motion compensation and turning velocity estimation; (c) Interferogram to estimate elevation angle of each point scatterer.}
    \label{fig:ISARimage_example}
\end{figure}
If the point scatterer is moving at a radial velocity $v_b$ from an initial position, $r_{b_0}$, and if the transmitter distance is $r_{tx}$, then the time-delays at the first and second elements are \par\noindent\small
\begin{align}
\tpone = \frac{r_{tx}+r_{b_0}+2v_bt}{c}\approx \frac{2r_{b_1}+2v_bt}{c} \\
\tp2 \approx \frac{2r_{b_0}+d\sin\theta_b+2v_bt}{c},
\end{align} \normalsize
where $c$ is the speed of light and $\theta_b$ is the elevation of the point target with respect to the radar when far-field assumption has been invoked. Note that in the case of automotive targets, the height of the point scatterer on the target is very low compared to the range and hence $\theta_b$ is likely to be very low. Also, if the transmitter and receiver are placed very close together (quasi-monostatic), then $r_{tx}\approx r_{b_0}$. Hence, the received signal at the $i^{th}$ radar receiver after down-conversion, is the superposition of returns from $B$ individual scatterers, if we ignore multiple bounces, as shown in \par\noindent\small
\begin{align}
\label{eq:RxSig}
\nonumber
\mathbf{S}_{{rx}_i}(t,\tau) = \sum_{b=1}^{B}a_b\text{rect}\left(\frac{\tau-\tpo}{T_{PRI}}\right)e^{-j2\pi f_c\tpo}\\e^{j\pi \beta (\tau-\tpo)^2} + \eta.
\end{align} \normalsize
Here, $\eta$ represents the noise and $a_b$ indicates the reflectivity of the $b^{th}$ scatterer as well as the two-way propagation factor including the gains of the transmitting and receiving antenna elements and channel effects. The received signal is represented as a 2D signal as fast time samples ($\tau$) are gathered across multiple pulse intervals constituting the slow time axis $t$. The total duration of the slow time interval is the coherent processing interval (CPI) and the CPI intervals are indexed at $k = 1\cdots K$. \\
\emph{Motion Compensation:}
In order to obtain ISAR images, we have to compensate for the translational motion of the target. Based on Fig.\ref{fig:JSTSP_SARsystem}, we obtain the  position coordinate estimate of the centroid of the target vehicle over the past $k-1$ CPI intervals based on which we estimate the starting range ($r_{0_k}$) and radial velocity, ($v_{r_k}$) for every $k^{th}$ CPI. Then assuming a first-order kinematic motion model, $r_k(t)=r_{0_k}+v_{r_k}t$ where $t$ spans from 0 to CPI. We use this to perform the coarse translation motion compensation as shown in \par\noindent\small
\begin{align}
\label{eq:MotionCompensation}
    \mathbf{\hat{S}}_{rx_{i}}(t,\tau) = \mathbf{S}_{rx_{i}}(t,\tau) e^{j2\pi\frac{2r_k(t)}{\lambda}}.
\end{align} \normalsize
The complexity of this centroid tracking approach is quite low since it involves the multiplication of each slow time column of the matrix with a one-dimensional vector. It is, therefore, preferred to other techniques proposed in literature such as the cross-correlation method \cite{chen1980target} which involves performing a cross-correlation between each range profile with a reference range profile (complexity $\mathcal{O}(n^3)$); and minimum entropy method which involves a 2D exhaustive search for the range and velocity parameters for carrying out the operation in \eqref{eq:MotionCompensation} \cite{wang1997minimum}.\\ 
\emph{Stretch Processing}
Next, we process the motion compensated received signal across the fast time through stretch processing. We consider a reference signal across the fast time shown by \par\noindent\small
\begin{gather}
\mathbf{s}_{ref}(\tau) = \text{rect}\left(\frac{\tau}{T_{ref}}\right)e^{j\pi \beta (\tau-\tau_{ref})^2}, \tau_{ref} = \frac{2R_{ref}}{c}.
\end{gather} \normalsize
Here $\tau_0$ is the time delay to a reference range position corresponding to the center of the range axis, and $T_{ref}$ is a pulse duration that is greater than $T_{PRI}$. An output 2D matrix $\mathbf{S}_{{out_i}}(t,\tau)$ is generated by element-wise multiplication of every $t^{th}$ row of $\mathbf{\hat{S}}_{{rx_i}}(t,\tau)$ with the conjugate of the reference signal as shown in \par\noindent\small
\begin{gather}
\label{eq:stretch}
\mathbf{S}_{{out}_i}(t,\tau) = \mathbf{\hat{S}}_{{rx}_i}(t,\tau) \mathbf{s}_{ref}^*(\tau)\\\nonumber
=\sum_{b=1}^{B}\text{rect}\left(\frac{\tau-\tpo}{T_{PRI}}\right) a_b e^{j\pi \beta \tau_{ref}^2} e^{-j\pi \beta \tpo^2}\\ e^{-j2\pi \frac{2r_{b_i}}{\lambda}}  
e^{-j2\pi f_{D_b}t}e^{-j2\pi \beta(\tau_{ref}-\tpo)\tau}.
\end{gather} \normalsize
Note that the operations in \eqref{eq:MotionCompensation} and \eqref{eq:stretch} can be combined into one without increase in complexity.  
In the resulting expression in \eqref{eq:stretch}, the first term within the exponent is constant with respect to $\tau, t$  and the receiver channel $i$ and absorbed into $a_b$. The second term has a phase term that is independent of $\tau$ and $t$ and changes very slightly with respect to $i$ due to the square of $c$ in the denominator. Hence, this term is ignored without loss of accuracy. The remaining three terms are linear functions of $i$, $\tau$ and $t$. After motion compensation, the Doppler frequency shift in \eqref{eq:MotionCompensation} arises entirely from the rotational motion of the target, and $(2\omega/\lambda)\rho_b = f_{D_b}$ where $\rho_b$ is the cross-range position of the $b^{th}$ point scatterer. Then, \par\noindent\small  
\begin{gather}
\nonumber
\mathbf{S}_{{out}_i}(t,\tau) =\sum_{b=1}^{B}\text{rect}\left(\frac{\tau-\tpi}{T_{PRI}}\right) a_b e^{-j2\pi \frac{2r_{b_i}}{\lambda}} \\
e^{-j2\pi (2\omega/\lambda) \rho_bt}e^{-j2\pi (2\beta/c) r'_{b_i}\tau},
\end{gather} \normalsize
where $2r'_{b_i}/c=\tau_{ref}-\tpo$.
By performing discrete Fourier transform (DFT) across the fast and slow time data, we obtain the range-crossrange plot $\chi$ for each $k^{th}$ CPI of the moving target as shown in \par\noindent\small
\begin{gather}
\nonumber
\chi_i(\rho,r) = \mathcal{DFT}\left\{\mathbf{S}_{{out}_i}(t,\tau)\right\}\\
\nonumber
= T_{PRI}T_{CPI} \sum_{b=1}^{B}a_b e^{-j2\pi \frac{2r_{b_i}}{\lambda}} \\\text{sinc}\left(\pi (2\omega/\lambda) (\rho-\rho_{b})T_{CPI}\right)
\text{sinc}\left(\pi (2\beta/c)(r-r'_{b_i})T_{PRI}\right),
\end{gather} \normalsize
Each point scatterer $b$ is convolved with a double-sided \emph{sinc} function and centered at the corresponding range and crossrange of the point scatterer in the two-dimensional plane. The duration of the fast and slow time axes are factored into the amplitude of the two-dimensional response. Note that, in the absence of motion compensation, the range-Doppler images provide us measurement estimates of the target centroid's range and Doppler at every CPI, which are subsequently fused with camera data and used for target state estimation. 
Second, we have considered a small angle approximation while carrying out the ISAR imaging because the vehicle undergoes small angular rotation (below $1^{\circ}$) during the short CPI that allow us to directly process the raw data with 2D DFT (of $\mathcal{O}(N^2)$ complexity). Alternatively, when the vehicles undergo large turns resulting in wide-angle synthetic apertures, direct two-dimensional integration through numerical techniques (of complexity $\mathcal{O}(N^4)$) may be considered. Or through nearest neighbors, the backscattered field data that correspond to polar data in spatial frequency may be reformatted onto a rectangular grid so that the DFT can be applied for the fast formation of the ISAR image. This introduces additional complexity of $\mathcal{O}(N^2)$ \cite{743822}.

In order to estimate the elevation of each point scatterer, we perform the very low complexity differential interferometry between the range-crossrange plots from the two channels. Specifically, we compute the phase difference of each point in the range-crossrange bin as shown in \par\noindent\small
\begin{gather}
\Theta(\rho,r) = \text{asin}\left(\frac{\lambda\left(\angle{\chi_2}(\rho,r)-\angle{\chi_1}(\rho,r)\right)}{2\pi d}\right),
\end{gather} \normalsize
to compute the interferogram $\Theta(\rho,r)$. Note that for accurate interferometry operation, the phase in each interferogram must be carefully unwrapped. This method of computing the elevation of radar point targets is effective when the point scatterers are well resolved in the range-crossrange space. In other words, there exists only a single point scatterer at any range-crossrange bin of $\chi_i$. However, note that the received signal in \eqref{eq:RxSig} will also consist of noise. Hence, the range-crossrange image and the corresponding interferogram will show noise artifacts. Additionally, in real-world scenarios, reflections from clutter and multipath are also observed in the radar signatures. 

To explain the radar signal processing, we provide an illustrative example of a simulated ISAR image of a car in Fig.\ref{fig:ISARimage_example}. Here, we consider a sedan of size $[5.7m \times 2.7m \times 2m]$ composed of over 6000 point scatterers corresponding to the centroids of the facets of the body of the car. The target undertakes a left turn before the radar with a dual channel receiver. In \cite{pandey2022classification}, a detailed description and open source model of the wideband electromagnetic target scattering was provided. We augment the single channel data to two receiver channels and process the radar data to obtain the ISAR image for a single frame in Fig.\ref{fig:ISARimage_example}b and the corresponding interferogram in Fig.\ref{fig:ISARimage_example}c. Since this is a simulation model, the trajectory followed by the target is known and hence we implement perfect translational motion compensation. The figure clearly shows the dimensions of the target along the range and cross-range dimensions. In this synthetic data, the shadowing of parts of the car by other parts is not considered. Hence, we are able to see the entire top-view cross-section of the vehicular target and estimate the size of the car (length and width). The wheels of the car undergo rotational motion and cause Doppler spread resulting in cross-range spread in the image. Note that due to the nature of the automotive vehicle target, the elevation angles are very low but sufficient to estimate the height of the car (based on the product of the largest elevation angle and range). However the interferogram image is particularly sensitive to noise compared to the ISAR image. This implies that the noise variance for the elevation estimate is greater than that of the range and crossrange. \\
\emph{Radar Detections:} We apply the ordered-statistic constant false alarm rate detector on the range-Doppler signature to dynamically identify the threshold based on the noise floor and identify radar detections in the image \cite{shor1991performances}. The computational complexity of this detector is based on the sorting algorithm and is higher than alternatives such as the cell-averaging CFAR (CA-CFAR) \cite{weiss1982analysis}. However, the OS-CFAR is widely used by radar practitioners due to its performance in dynamic noise and clutter conditions. The high-range resolution imaging described in this section results in multiple detections for each target. The detections that belong to a single target are clustered together from which the range, Doppler frequency and elevation of the target's centroid is identified.
\subsection{Camera Data Processing}
The main objective of the camera data processing is to detect a known camera target vehicle in the image and convert its centroid position in the camera's image to 3D radar coordinate frame. \\
\emph{Camera Calibration:} Consider a
monocular digital camera image where each pixel is indexed by $\mathbf{u} = [p_u,p_v]^T, p_u = 1 \cdots U, p_v = 1 \cdots V$. Here $p_u$ indicates the vertical or longitudinal position aligned to $z^{\C}$ axis and $p_v$ indicates the lateral or horizontal position aligned to the $y^{\C}$ axis while the $x^{\C}$ axis is the optical axis.
Camera calibration is the process of learning two sets of parameters using a linear camera pinhole model. They are the \emph{extrinsic} camera parameters that facilitate the transformation of a 3D point, $\xw$ in $\R$, to a 3D point, $\xc$, in $\C$ and \emph{intrinsic} parameters that enable perspective projection of $\xc$ to a 2D point $\mathbf{u}$ in the camera's image plane. The intrinsic parameters include the focal length of the camera lens ($f$), the pixel densities along the two image dimensions ($m_u,m_v$), and the position of the principal point ($p_{u_0},p_{v_0}$) through which the optical axis emanates. Any point $\xc$ is projected to \par\noindent\small
\begin{align}
\label{eq:CamUV}
p_u = m_uf\frac{y^c}{x^c}+p_{u_0}\\
p_v = m_vf\frac{z^c}{x^c}+p_{v_0}.
\end{align} \normalsize
The basis of intrinsic camera calibration is that the homogeneous representation of the image projected point $[\mathbf{u},1]^T$ is equivalent to the 3D point $\bar{\mathbf{u}} = [x^c\;x^cp_u\;x^cp_v]^T$. Thus, \par\noindent\small
\begin{align}
    \begin{bmatrix}
    \mathbf{u} \\
    1
    \end{bmatrix}\equiv\bar{\mathbf{u}} = \begin{bmatrix} 
                        1 & 0 & 0 & 0\\
                        p_{u_0} & m_uf & 0  & 0\\
                        p_{v_0} & 0 & m_vf & 0
                        \end{bmatrix}
                        \begin{bmatrix}\xc \\ 1\end{bmatrix} = \mathbf{M}^{int}_{3 \times 4}\xch,
\end{align} \normalsize
where $\mathbf{M}^{int}_{3 \times 4}$ is the intrinsic calibration matrix that must be estimated and $\xch$ is the homogeneous representation of $\xc$. The extrinsic parameters enable the alignment of $\C$ with $\R$ through the extrinsic matrix $\mathbf{M}^{ext}_{4 \times 4}$. This comprises the Euler rotation matrix $\textbf{R}^{\C\R}$ that aligns the axes between the two coordinate frames; and the translation vector $\mathbf{t}^{\C\R}$ between the origins of the two coordinate frames. Thus, \par\noindent\small
\begin{align}
\xch = \begin{bmatrix}
\mathbf{R}^{\C\R}_{3\times 3} & \mathbf{t}^{\C\R}\\
\mathbf{0}_{3\times 1} & 1
\end{bmatrix}
\xwh = \mathbf{M}^{ext}_{4 \times 4}\xwh.
\end{align} \normalsize
The intrinsic and extrinsic matrices are combined to form the projection matrix $\mathbf{P}^{cal}_{4 \times 4}$ which converts a 3D point $\mathbf{x}$ to 2D $\mathbf{u}$ through \par\noindent\small
\begin{align}
  \begin{bmatrix}
  \mathbf{u}\\
  1
  \end{bmatrix} \equiv \bar{\mathbf{u}} = \mathbf{M}^{int}_{3 \times 4}\mathbf{M}^{ext}_{4 \times 4}\xwh=\mathbf{P}^{cal}_{4 \times 4} \xwh.
\end{align} \normalsize
In this work, both the intrinsic and extrinsic parameters are estimated using the popular open source calibration algorithm involving a checkerboard patterned target that is viewed from the camera from two different orientations \cite{zhang2000flexible}. We have restricted the above discussion to conventional cameras. If wide-angle fish-eye cameras are to be used instead of the conventional cameras, additional calibration steps are required to resolve radial distortion errors \cite{kannala2006generic}. \\
\emph{Object Detection}
Detection of an object (vehicle or pedestrian) within a camera's image has been tackled using both traditional machine learning \cite{viola2001rapid} and deep learning techniques using convolutional neural networks and their variants \cite{girshick2014rich,girshick2015fast,ren2015faster}. The main challenge in these algorithms is to handle the multiple resolutions and orientations of an object in the image plane. Hence the object detection methods essentially trade-off between detection accuracy specified through the number of false positives per image and the computational cost for capturing the multi-resolution image features. 
\begin{figure}
    \centering
    \includegraphics[scale=0.23]{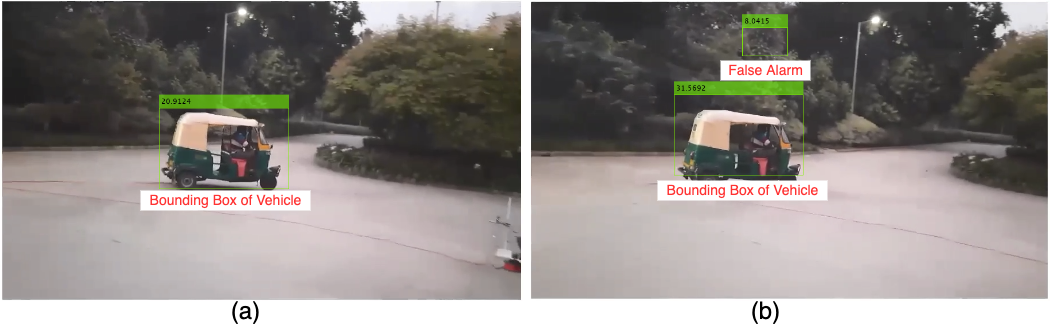}
    \caption{Results from object (auto-rickshaw) detection based on \cite{dollar2014fast} (a) without false alarm, (b) with false alarm.}
    \label{fig:CarBoundingBox}
\end{figure}
For example, in Fig.\ref{fig:CarBoundingBox}, the objective is to detect and segment the vehicle (auto-rickshaw) in the images as the vehicle moves and turns. In Fig.\ref{fig:CarBoundingBox}a, the object was correctly detected and segmented. However, in Fig.\ref{fig:CarBoundingBox}b an additional object was also detected which corresponds to a false positive. We use open source machine learning-based object detection code provided with \cite{dollar2014fast}. The key insight of this algorithm is that it exploits the fractal statistics in natural images, termed internal channel features, to reliably predict image structures across multiple resolution scales which are termed aggregate channel features. Since the image features are obtained through extrapolation rather than explicit prediction, the overall computational complexity is considerably reduced. The performance of the algorithm has been shown to be very effective for automotive target detection (pedestrians and vehicles). 
\subsection{Fusion of Camera and Radar Data}
Once the camera projection matrix is learned it will be possible to estimate $\ye,\ze$ for a center point of a target $t$ in the image plane. However, since the transformation operation is inherently lossy, we cannot recover the depth $\xe$ (depth) from the image point $\mathbf{u}_k$ at every $k$. In other words, any number of 3D scenes can result in the same 2D image. Depth estimation from a single camera image is considerably more challenging than stereo camera images and several approaches have been proposed in literature including the use of machine learning algorithms \cite{saxena2005learning,liu2015learning,cao2016exploiting}. In this work, we supplement the camera data with radar data to track the object.

We model the automotive target motion using the curvilinear state model based on constant turn rate and velocity (CTRV) as described by \cite{4632283,8916654}. 
Here, the state vector of the vehicle at any time instant $k$ is given by $\mathbf{x}_k=[\xe;\ye;\vx;\vy;\omega_k]^T$ where $(\vx,\vy)$ describe the 2D velocity in Cartesian coordinates. Note that we also obtain the $\ze$ from the longitudinal position estimates from the camera image and the elevation estimates from radar interferometry. However, we do not use those measurements since they do not have a bearing on the computation of the turning velocity and very weakly impact the translation motion compensation. They are useful, however, for correct data association from the two sensors in scenarios where there are multiple targets.
The state at $k+1$ time step is predicted based on \par\noindent\small
\begin{align}
\label{eq:KalmanState}
\mathbf{x}_{k+1} = \mathbf{f}\left(\mathbf{x}_{k}\right) + \mathbf{G}\left(\mathbf{x}_k\right)\mathbf{w}_k,
\end{align} \normalsize
where $\mathbf{f}(\cdot)$ is a vector-valued function that models the state transition and is given by \par\noindent\small
\begin{align}
\label{eq:StateUpdate}
\mathbf{f}(\mathbf{x}_{k}) = \left[\begin{array}{c}
x_k + \frac{v_{x_k}}{\omega_k}\sin(\omega_k T) -\frac{v_{y_k}}{\omega_k}(1-\cos(\omega_k T)) \\
y_k + \frac{v_{x_k}}{\omega_k}(1-\cos(\omega_k T)) +\frac{v_{y_k}}{\omega_k}\sin(\omega_k T) \\
v_{x_k}\cos(\omega_k T) - v_{y_k}\sin(\omega_k T) \\
v_{x_k}\sin(\omega_k T) + v_{y_k}\cos(\omega_k T) \\
\omega_k
\end{array}\right],
\end{align} \normalsize
and $T$ is the time interval between updates (CPI).
$\mathbf{G}(\mathbf{x}_k)\mathbf{w}_k$ in \eqref{eq:KalmanState} is the process noise matrix with $\mathbf{Q}$ process noise covariance matrix. In a CTRV model, the speed and the turn rate are assumed to be nearly constant and we model the acceleration and rate of change of yaw rate (which form vector $\mathbf{w}_k$) as noise processes with zero mean and $\sigma_a$ and $\sigma_{\alpha}$ standard deviations respectively. $\mathbf{G}$ models the relationship between the target state and $\mathbf{w}_k$ and is given by \par\noindent\small
\begin{align}
\label{eq:Gmat}
\mathbf{G} = \left[\begin{array}{cc}
\frac{T^2}{2} \cos(\omega_k T) & 0 \\
\frac{T^2}{2} \sin(\omega_k T) & 0 \\
T \cos(\omega_k T) & 0 \\
T \sin(\omega_k T) & 0 \\
0 & T
\end{array}\right].
\end{align} \normalsize
The target space is updated based on measurements vector, $\mathbf{z}_k = [r_k\;f_{D_k}\;p_{v_k}]^T$, provided by the automotive radar and the camera. The measurements are mapped to the state space through \par\noindent\small
\begin{align}
\mathbf{z}_k = \mathbf{h}(\mathbf{x}_k) + \mathbf{\nu}_k,
\end{align} \normalsize
where $\mathbf{h}(\cdot)$ is the function between the measurements and state space and $\mathbf{\nu}_k$ is the observation noise vector with $\mathbf{R}$ measurement noise covariance matrix. 
The radar provides range and Doppler frequency measurements, $[r_k,f_{D_k}]$ which are non-linear functions of the position and velocity of the target as shown in \par\noindent\small
\begin{align}
\label{eq:MeasKalman}
r_k = \sqrt{x_k^2+y_k^2}\\
f_{D_k} = \frac{2f_c}{c} \left(\frac{2x_kv_{x_k}}{r_k}+\frac{2y_kv_{y_k}}{r_k}\right).
\end{align} \normalsize
The corresponding observation noise for the radar range and Doppler frequency estimates are governed by the radar bandwidth and CPI respectively. They are modeled, therefore, as uncorrelated random variables with zero mean and standard deviations which are directly proportional to the corresponding resolutions. The camera coordinate, $p_{v_k}$, depends on the focal length and the pixel resolution of the camera as described in Section III.b (equations (14) to (16)) while the standard deviation assumed for the process noise is assumed to be half the width of the smallest bounding box required for object detection.   
Due to the non-linear state space prediction equations and their measurement equations, we used the extended Kalman filter framework to estimate the state space \cite{blackman1999design}. The EKF is a nonlinear estimation algorithm based on consecutive linearization of the equations carried out via first-order Taylor series expansions. The EKF is preferred to other non-linear estimation methods such as the second order extended Kalman filter \cite{norgaard2000new}, unscented Kalman filter \cite{1271397}, and particle filter \cite{carpenter1999improved} due to its low computational complexity ($\mathcal{O}(n^3)$ versus $\mathcal{O}(n^4)$). Due to the approximations in the system equations, the EKF is known to generate error and instability conditions for more complex target motions. However, within the limited scope of our problem - a single vehicle undergoing turns - the EKF performs satisfactorily. 
We summarize the EKF algorithm below. %Since these equations are standard and widely used in literature and practice and numerous good references are available for the interested reader, they are not included in the manuscript but are provided here for the reference of the reviewer.  
\begin{enumerate}
\item Initialize target state $x_{0/0}$ and error covariance $P_{0/0}$
\item For $k<K$, predict the current state, error covariance, $\mathbf{P}$, and Kalman gain, $\mathbf{K}$, from previous time estimates \par\noindent\small
\begin{align}
\mathbf{x}_{k/k-1} = \mathbf{f}(\mathbf{x}_{k-1/k-1})\\
\mathbf{P}_{k/k-1} = \mathbf{F}_{k-1}\mathbf{P}_{k-1/k-1}\mathbf{F}_{k-1}^T + \mathbf{Q}_{k-1}\\
\mathbf{K}_k = \mathbf{P}_{k/k-1}\mathbf{H}_{k}^T(\mathbf{H}\mathbf{P}_{k/k-1}\mathbf{H}_{k}^T+\mathbf{R}_k)^{-1},
\end{align} \normalsize
where $\mathbf{F}$ and $\mathbf{H}$ are the Jacobian matrices for the vector-valued functions $\mathbf{f}$ and $\mathbf{h}$ respectively with respect to the state.
\item Update the current state using current measurements using the Jacobian computed from the non-linear equation and update the corresponding error covariance \par\noindent\small
\begin{align}
\mathbf{x}_{k/k} = \mathbf{K}_k (\mathbf{z}_k - \mathbf{h}(\mathbf{x}_{k/k-1}))\\
\mathbf{P}_{k/k} = (1-\mathbf{K}_k\mathbf{H}_k)\mathbf{P}_{k/k-1}
\end{align} \normalsize
\item Repeat step 2 through 4 till the target trajectory is completed.
\end{enumerate}
The Kalman gain, $\mathbf{K}$, is a $[3 \times 5]$ matrix that controls the impact of the three sensor measurements (range, Doppler velocity from radar, and lateral position from camera) on the five-state target estimation (two-dimensional position, two-dimensional velocity, and yaw rate) at each time step. When the Kalman gain value $\mathbf{K}_{m,n}$ is high (close to 1), this implies that the $n^{th}$ state variable is strongly correlated to the $m^{th}$ measurement and that the measurement is reliable (compared to the predictions). On the other hand, when $\mathbf{K}_{m,n}$ is low (close to zero), this implies that the $n^{th}$ state variable is not correlated to the $m^{th}$ measurement and should be updated based on predictions from its prior state or from one of the other measurements. Usually, when there are no instances of sensor failure, the Kalman gain values start from a high value, when predictions are unreliable, and then converge to a lower value (say around 0.5) when the estimates are based on prior state predictions and current measurements. In the event, that one of the two sensors fails to provide a detection due to malfunction or because the target is outside its FoV, then the corresponding $\mathbf{K}_{m,n}$ is set as 0 and the measurements from the other sensor are used to correct the state predictions. For example, if the camera fails, then $\mathbf{K}_{3,n}=0$ for $n=1:5$ and $\mathbf{z}$ only consists of range and Doppler velocity measurements. In the case both sensors fail to produce measurements at any particular time step, then the states are updated based on prior predictions and not further corrected. Subsequently, the corrected target state is used for translation motion compensation of the range-Doppler ambiguity plot and for mapping the range-crossrange ISAR image at $k+1$. 
Note that in this work we have confined the problem to a single target in the FoV of the sensors. Hence, target detections (from either of the sensor) at timestamp $k+1$ that do not fall within a predefined neighborhood of the estimated position of the target are discarded. In this manner, false alarms of both radar and camera are reduced. In scenarios with multiple targets, joint probabilistic data association and multiple hypothesis testing will have to be incorporated within the Kalman framework. The discussion on multiple target tracking is outside the scope of this paper.
\section{Simulation Model and Results}
In this section, we present the simulation model of a camera and radar tracking a dynamic target undertaking turns. The data from the two sensors are fused and the target state is estimated which is subsequently used to generate the corresponding ISAR images. The entire simulation is carried out in MATLAB using the automated driving toolbox. 
\subsection{Experimental Set Up}
\emph{Road Geometry:}
We consider a four way flat road junction geometry as shown in Fig.\ref{fig:EKFResults_Sim}i.a-d. The north-south road, aligned along the X axis of $\R$ is perpendicular to the east-west road which is aligned to the Y axis. The height axis above the road thus corresponds to the Z axis. The origin is assumed at the lower south-west corner of the figure. Each road is a 78m long dual carriageway that allows traffic in both directions. The solid yellow lines in the figure indicate the road divider separating the bi-directional traffic flow in each road segment while the dashed yellow lines demarcate the lanes within each section. Each lane is 3.6m wide as per common road design standards. An additional lane is provided for onward traffic in order to facilitate right turns (south to east, east to north, and so forth). For example, the left section of the south road segment consists of three lanes while the right section consists of two lanes. So a vehicle that turns from south to east will move from the right-most lane of the left section of the south road to the lowest lane of the upper section of the east road. We assume line-of-sight conditions with no building or other forms of occlusions in the road layout. 

\emph{Sensor Specifications:}
We assume that the ego vehicle upon which the radar and camera sensors are mounted is not moving. We assume that the ego vehicle is a simple cuboid of $[4.7 \times 1.8 \times 1.4]m$ dimensions with its length oriented along the direction of the lane. Thus the yaw (angle of the vehicle with respect to the positive X axis) is zero. The center point of the ego vehicle is located in the south road segment at $[10,42.6,0.7]m$. Forward-looking radar and camera are mounted on the ego vehicle as indicated in Fig.\ref{fig:EKFResults_Sim}i.a-d. The camera is assumed to be located 0.1m below the top of the ego vehicle and at 0.7m of the wheelbase of the vehicle. The camera's intrinsic parameters are the following- the focal lengths along $p_u$ and $p_v$ are $[800,800]$, the image sizes  are ($[480,640]$) and the principal point is located at ($[320,240]$). The camera's yaw, pitch, and roll are all $0^{\circ}$. The toolbox's vision sensor is configured to detect targets up to a range of 100m with a detection probability of 90\% and a statistical probability of false positives per image of 10\%. The minimum object size in the image plane that can be detected is $[15 \times 15]$ pixels. \\
Next, we discuss the radar model. We assume that the radar is transmitting a linear frequency modulated wave as discussed in Section II. The carrier frequency is 77GHz and the chirp rate is $60\times 10^{12}Hz/sec$. The pulse repetition frequency of the radar is 40kHz resulting in a maximum unambiguous range of 40m and the range resolution is 0.1m. We assume that the CPI for the radar is 0.1s which also corresponds to the frame rate of the camera. In other words, the camera and radar data are time stamped every 0.1s. The radar is located 0.1m above the ground at the front overhang of the ego vehicle. The yaw, roll, and pitch of the radar are all zero. Therefore, in this simulation set up the axes of $\C$ and $\R$ are perfectly aligned though they have different origins. The radar is characterized by a false alarm rate of $1\times 10^{-6}$ and a detection probability of 90\%. The FoV of the radar is $120^{\circ}$ along the azimuth and $90^{\circ}$ along the elevation. Figure.\ref{fig:EKFResults_Sim}i.a-d shows the coverage area of the camera and the radar. Note that this figure is not drawn perfectly to scale and is provided to give a general notion of the coverage of the sensors. \\
\emph{Target Details:} We model the vehicle as a cuboid of dimensions $[4.7m \times 1.8m \times 1.4m]$ which is the size of a typical mid-size car. The wheelbase for this car is 2.8m with a rear overhang of 1m and the front overhang of 0.9m. The target is modeled as an extended target with 12 triangular faces/mesh (2 per side). The radar cross-section (RCS) of each mesh is calculated based on the analytical expression for RCS of metal triangle \cite{pandey2020database}. We consider four possible turn trajectories for the target as indicated in Fig.\ref{fig:EKFResults_Sim}a-d. 
In the first trajectory, the target starts in the south lane at $[20m, 39.5m, 0m]$, moves forward and performs a U-turn, and then returns to the south lane at $[20m, 35.4m, 0m]$. Thus the target is mostly within the FoV of both sensors. Next, we will consider the trajectory shown in Fig.\ref{fig:EKFResults_Sim}i.b. Here, the target starts in the north lane at $[58m, 39m, 0m]$ and then comes south-wards and performs a U-turn, and then returns to the north lane at $[58m, 42.6m, 0m]$. The target lies within the camera's FoV for almost the entire duration and is within the radar's FoV while it is performing the turn. The next trajectory that we consider is the target turning right from east to north as shown in Fig.\ref{fig:EKFResults_Sim}i.c from $[39m, 20m, 0m]$ to $[58m, 42.6m, 0m]$. Again, the target is initially outside the FoV of the camera but eventually enters it. The opposite is true with respect to the radar. The last trajectory is when the target turns right from the west ($[39m, 58m, 0m]$) to the south ($[20m, 35.4m,0m]$). The target is mostly within the FoV of both the sensors for the most part. 
We will refer to these four trajectories as SSUT, NNUT, ENRT, and WSRT henceforth. The duration of motion for each case is 6 seconds and the speed of the vehicle is $6m/s$.
\subsection{Simulation Results}
\begin{figure*}
\centering
\includegraphics[width=7in,height=4in]{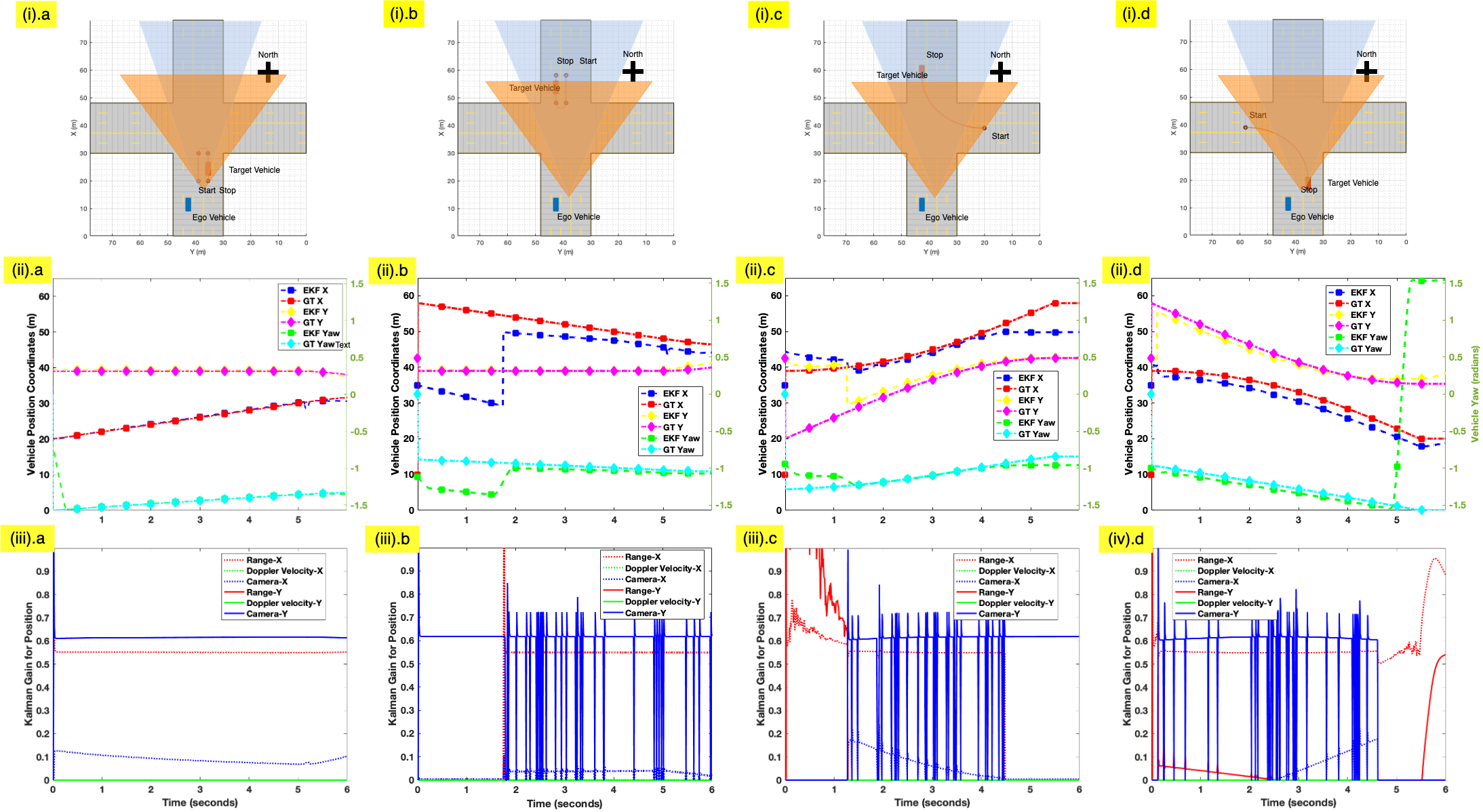}
\caption{Top row shows the simulation setup for radar-camera sensor fusion and target trajectories along (a) south lane to south lane U-turn (SSUT), (b) north lane to north lane U-turn (NNUT), (c) east lane to north lane right turn (ENRT), and (d) west lane to south lane right turn (WSRT). The second row shows the corresponding target trajectory estimates from sensor fusion for (e) SSUT, (f) NNUT, (g) ENRT, and (h) WSRT. The third row shows the Kalman gain for position estimates based on sensor measurements (radar range, Doppler velocity, and camera-based lateral position of object). }
\label{fig:EKFResults_Sim}
\end{figure*}

In this section, we discuss the effectiveness of sensor fusion for target tracking and ISAR imaging. We present the estimated position ($\xe, \ye$) and yaw rate ($\omega_k$) of the target based on the camera-radar detections for the four trajectories in the second row of Fig.\ref{fig:EKFResults_Sim}. 
The third row shows the Kalman gain over the duration of the trajectory. %In the case both sensors fail to produce measurements at any particular time step, then the states are updated based on prior predictions and not further corrected. 
From the results in the third row, it is evident that $\mathbf{K}_{1,1}$ shows a strong correlation between the radar range measurement and $\xe$ while $\mathbf{K}_{3,2}$ shows the strong correlation between the horizontal/lateral position of the target in the camera's image to $\ye$. Both of these measurements are not directly correlated to the Doppler velocity and hence $\mathbf{K}_{2,1}$ and $\mathbf{K}_{2,2}$ are zero throughout the duration of the trajectory. 

{SSUT:} Since the target falls within the FoV of both the sensors for most of the duration of the trajectory the Kalman gain plot is fairly stable in Fig.\ref{fig:EKFResults_Sim}iii.a. The target is also close to the sensors. This results in strong SNR in the case of radar and a large object for detection by the camera. Due to all of these factors, the estimation of the target state is fairly accurate in this trajectory. 

{NNUT:} Here, the target object is farther from both the sensors. First, it is outside the FoV of the radar, which is reflected in the zero values of $\mathbf{K}_{1,1}$ and $\mathbf{K}_{1,2}$ and results in large errors in the early estimates of $\xe$. Later, even when it is within the radar FoV, there is a discrepancy of approximately $2m$ between $\xe$ and the ground truth. This is because the camera image and radar locate the centroid of the vehicle at the midpoint of its front end/rear end (whichever is facing the sensors) rather than the center of the vehicle. 
There are also instants when the camera does not detect the object due to smaller perspective spans in the image. This effect is noticeable in the pattern of $\mathbf{K}_{3,2}$. However, it does not cause a large drop in the accuracy of the camera estimates. 

{ENRT:} Here, both $\xe$ and $\ye$ are inaccurately estimated at the beginning when the vehicle is outside of the FoV of both sensors. This is also reflected in the Kalman gain plot in Fig.\ref{fig:EKFResults_Sim}iii.c. Later, the results improve while the target remains in the FoV of both sensors. Towards the end of the trajectory, the vehicle is outside the radar FoV which causes a deterioration in the estimate of $\xe$. Similar observations can be made for the last case corresponding to WSRT. In all four cases, we observe that the yaw rate estimate is reasonably accurate when the target is within the FoV of both sensors. In the above discussion, we have assumed that both the sensors are performing with high detection metrics (above 90\%). In the appendix, we discuss the sensor fusion performance when the detection metrics of both sensors are poor due to either environmental conditions or instrument defects.

\begin{figure*}
    \centering
    \includegraphics[width=7in,height=3.5in]{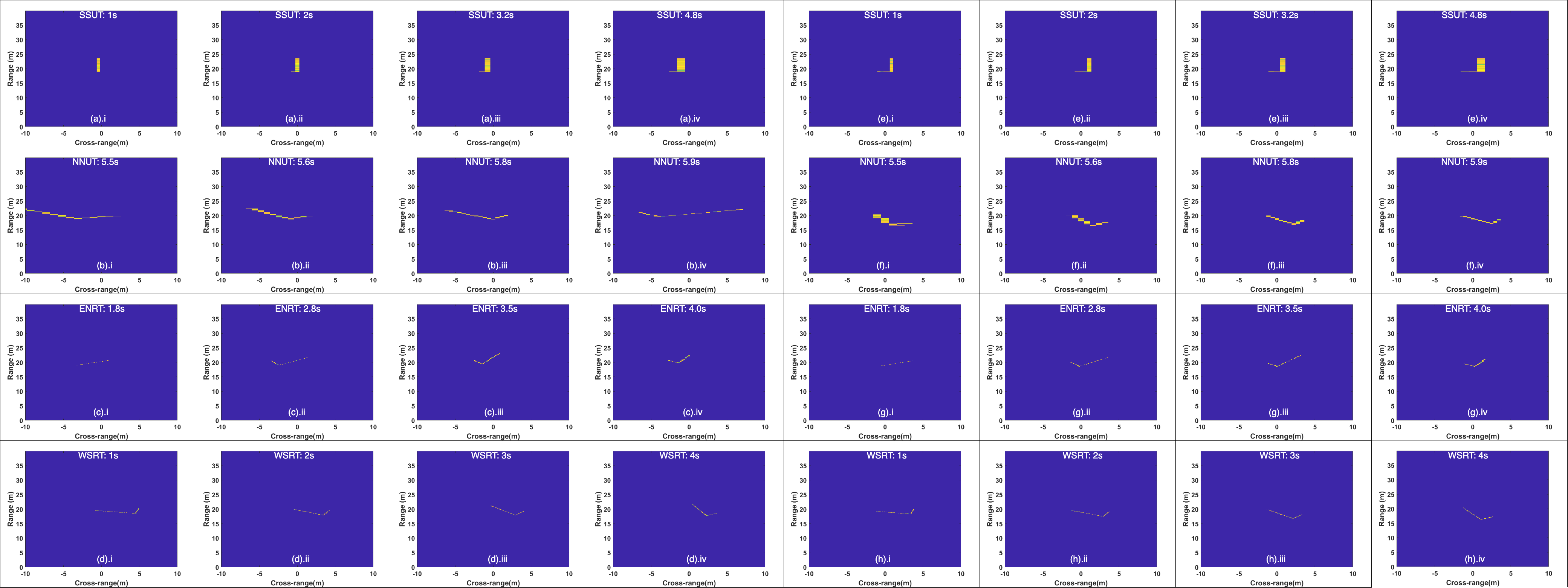}
    \caption{ISAR images from simulated data with motion compensation based on: (i-iv) target state estimation from camera-radar fusion; and (v-vi) ground truth target state. Rows correspond to (a) south to south U turn (SSUT), (b) north to north U-turn (NNUT), (c) east to north right turn (ENRT) and (d) west to south right turn (WSRT).}
    \label{fig:SimCAM_GTImages}
\end{figure*}

Next, we discuss the results obtained for ISAR imaging of the vehicle. The results for all four trajectories are presented in Fig.\ref{fig:SimCAM_GTImages}. 
The ISAR images are generated after motion compensation using turning velocity estimates obtained from the radar-camera fused data. These are presented in the first four columns (left four) corresponding to Fig.\ref{fig:SimCAM_GTImages}a-d. These images are then compared with those obtained from ground truth information of the target trajectory. These are presented in the last four columns of each row (right four) corresponding to Fig.\ref{fig:SimCAM_GTImages}e-h. All the figures span 40m along range and 20m along crossrange and are mirrored along the crossrange to keep positive sense of the Y axis. Note that we may not be able to generate images even with the perfect knowledge of the ground truth target information. If the vehicle undergoes very little yaw with respect to the ego vehicle upon which the sensor is mounted, then the turning velocity is very low resulting in extremely high cross-range resolutions. 
\begin{table}[tb]
    \centering
    \begin{tabular}{p{2cm}|c|c|c}
        \hline\hline
        Trajectory & GT images (\#) & Fused images (\#) & SSIM (\%) \\
         \hline \hline
         South to South U Turn(SSUT) & 56 & 40 & 97.4  \\
         North to North U Turn(NNUT) & 6 & 15 & 85.7  \\
         West to South Right Turn(WSRT) & 54 & 46 & 97.5  \\
         East to North Right Turn(ENRT) & 52 & 40 & 97.5  \\
         \hline \hline
    \end{tabular}
    \caption{Comparison of fused ISAR images with ground truth (GT) images.}
    \label{tab:ISARSimResults}
\end{table}
For example, in our problem, if the turning velocity is below $0.01 rad/s$, then the resulting crossrange resolution is $1.98m$ which is greater than the width of the vehicle and hence not useful for imaging. Hence, we do not generate images for very low values of turning velocity. The images obtained from the sensor fused data are compared with the ground truth data through the popular image comparison metric called the structural similarity index (SSIM) \cite{hore2010image}. The summary of the ISAR image results is presented in Table.\ref{tab:ISARSimResults}.

We first consider the results from SSUT trajectory along the first row. Here, due to the orientation of the vehicle with respect to the ego vehicle as shown in Fig.\ref{fig:EKFResults_Sim}a, we see only one long side and the rear end of the vehicle at different time instants of the target trajectory. The other two sides of the vehicle are occluded from the sensors. As the target moves along a straight line to the right of the vehicle, it still presents a variation in the target aspect resulting in a change in the yaw. Hence, we are able to generate 56 ISAR images out of the total 60 possible intervals in the duration of the vehicular motion. For higher turning velocity, the crossrange resolution is finer and that is reflected in the thickness of the pixels across the different time instants. With the sensor fused data, we are able to generate 40 ISAR images because there are some time instants where the predicted turning velocity becomes very low. However, these images are very similar for the most part to the ground truth images as evidenced by the high SSIM score.

Next we discuss the results from NNUT trajectory. In this trajectory, the target object is directly before the ego vehicle. Hence, for most of its trajectory, the yaw with respect to the sensors is zero. Hence, the turning velocity is infinite and we cannot generate ISAR images. Out of the 60 time intervals, it is possible to generate images for only 6 intervals when the vehicle is actually undergoing the turn. The number of ISAR images generated from the sensor fused data are higher in this case because of higher (incorrectly) estimated values of the turning velocity. Secondly, as mentioned earlier, there is some error in both sensor estimates since we do not correctly estimate the position of the center of the vehicle and instead track the center of the front side of the vehicle. Due to these reasons, we observe that some of the ISAR images generated from the sensor fused data show wide spread across the crossrange, especially in comparison to the ground truth. This is reflected in the lower mean SSIM score reported in Table.\ref{tab:ISARSimResults}. However, we are still able to observe (for the most part) the size and shape of the vehicle.

Next we consider the two right turn trajectories - ENRT and WSRT. In both cases, the turning velocity from ground truth and the estimate are fairly high for most of the duration of the vehicle and we are able to generate a large number of ISAR images. Interestingly, the number of images from ground truth and sensor fusion are exactly equal in both cases because the estimation of the target state along these two trajectories is fairly accurate as shown in Fig.\ref{fig:EKFResults_Sim}. Due to occlusion, different sides of the target cuboid are visible to the sensors for each case. Based on how the orientation of these images change, we can make inferences regarding the trajectory of the target. Due to the close agreement between the sensor fused images and ground truth images, we obtain high mean SSIM scores as reported in Table.\ref{tab:ISARSimResults}
\section{Measurement Setup and Results}
In this section, we present the measurement setup to collect experimental data from Texas Instruments AWR-1843 77GHz millimeter wave radar and a smartphone camera of a target vehicle as described in \cite{pandey2022classification}. The radar parameters are configured as per Table.\ref{tab:RadarParameters}.
\begin{table}[htp]
\caption{Automotive radar parameters for generating ISAR images}
    \centering
    \begin{tabular}{c|c}
    \hline \hline
        Parameters & Values  \\
    \hline \hline
         Carrier frequency ($f_c$) & 77GHz \\ 
       Range resolution  & 0.075m \\
       Maximum unambiguous range & 100m \\
       Radar bandwidth & 2GHz \\
        Chirp duration ($T_{PRI}$) & 400$\mu$s\\
        Coherent processing interval ($T_{CPI}$) & 0.1s\\
        Doppler resolution & 10 Hz \\
        Minimum cross-range resolution & 0.19 m \\
        Transmitted power ($P_t$) & 25dBm \\ 
    \hline \hline
    \end{tabular}
    \vspace{-2mm}
    \label{tab:RadarParameters}
\end{table}
In the measurement setup, the radar is mounted on a static platform which is assumed to be located at the origin of $\R$ frames. The radar images are generated for every CPI interval of 0.1 seconds. Here, the radar consists of only a single channel and hence we do not obtain estimates of the height of the vehicle. 
A smartphone camera of $1920 \times 1080$ pixel size is mounted at $[2m,-2m,0.5m]$ and is oriented with yaw of $-45^{\circ}$ with respect to the radar axis to ensure that the target object remains within the camera FoV during the entire duration. The camera frame rate is 30 frames per second and is time stamped. 
\begin{figure}
    \centering
    \includegraphics[scale=0.25]{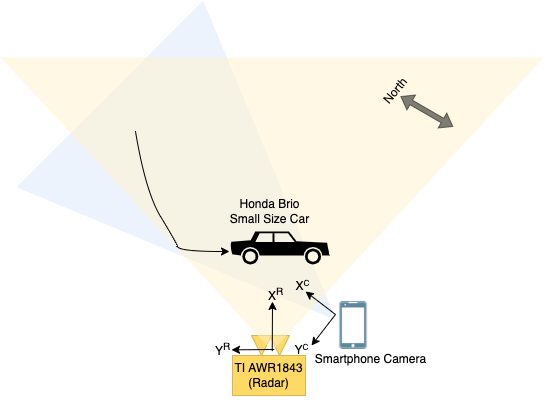}
    \caption{Measurement setup of camera radar data collection of a small size car executing a left turn}
    \label{fig:MeasSetup}
\end{figure}
We select the images corresponding to the 0.1 second interval. Hence, unlike the simulation data, the measurement data from the two sensors are not perfectly synchronized. The camera is calibrated to the radar's coordinate frame based on the methods discussed earlier. We consider a Honda Brio car which is a small size car of $[3.6m \times 1.68m \times 1.5m]$ dimensions. The car undertakes a left turn from the north to the east in front of the radar as shown in Fig.\ref{fig:MeasSetup}.

\begin{figure*}
    \centering
    \includegraphics[width=7in,height=3.89in]{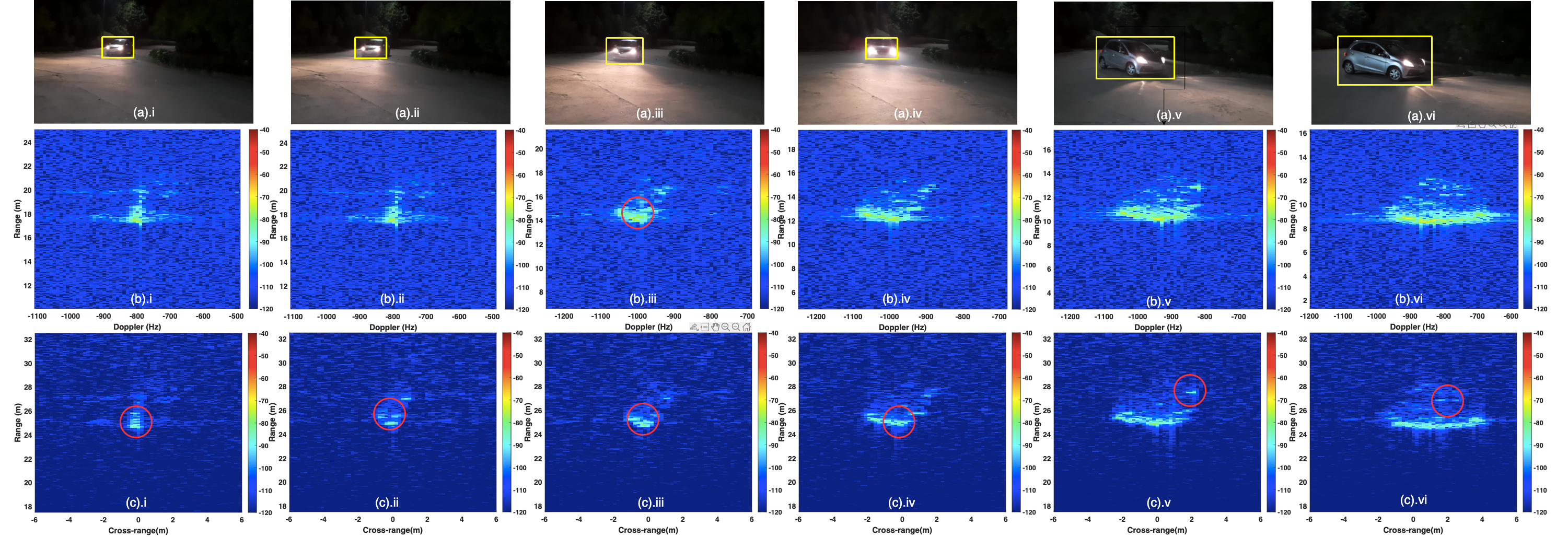}
    \caption{(a)i-vi: Camera images of a small size car executing a left turn from north to the east before a radar. (b)i-vi: Range-Doppler plots generated from two-dimensional Fourier transform of slow-time fast-time radar measured data from TI AWR-1843. (c) ISAR images were obtained after target state estimation from camera-radar fused data.}
    \label{fig:MeasCAM_GTImages}
\end{figure*}

We present the measurement results for six frames in Fig.\ref{fig:MeasCAM_GTImages}. 
The top row shows the camera images with the object within the bounding box. Due to the darker background, we applied lightening filters to improve the effectiveness of object detection. Based on the camera estimates and the radar estimates, we track the trajectory of the centroid of the target. The second row of Fig.\ref{fig:MeasCAM_GTImages}(b)i-vi shows the range-Doppler plots obtained from the radar without any type of motion compensation. In order to improve the clarity of the images, we have cropped the range-Doppler plot to show the region within $\pm 6m$ and $\pm 300Hz$ of the range and Doppler, respectively, corresponding to the peak scatterer on the vehicle. The axes of these images change based on the range and radial velocity of the vehicle with respect to the radar. Using the translational motion compensation steps outlined earlier, we generate the corresponding ISAR images shown in the third row. Additionally, CFAR algorithm is also implemented on the raw data to improve the noise performance. The crossrange axis is obtained from the estimate of the turning velocity.
From the ISAR images, we are clearly able to observe the orientation of the vehicle change from width-wise (when it is approaching the radar) in (c)i-iii to length-wise (when it is parallel to the radar Y axis) in (c)iv-vi. One or more side edges of the vehicle can be observed in each of the images while the remaining edges are occluded. With these images, we able to obtain a fairly good estimate of the top-view (bird's eye view) dimensions of the car. Unlike the simulation results presented in Fig.\ref{fig:ISARimage_example}, all four edges are not visible. Also, the micro-Dopplers from the wheels are very weak in this car. 

We observe that the ISAR images are more focused compared to the range-Doppler plots due to the motion compensation. Specifically, compare the regions red circled in (c)i-iii with the corresponding regions in (b)i-iii. The point scatterers are better defined in the ISAR images than in the range-Doppler plots. Across the ISAR images, the crossrange resolution changes with the turning velocity. This is reflected in the slight change in the cell sizes in the images. Even when the car is parallel to the radar Y axis, we observe some features from the back of the car possibly due to the transparent windows. The figures demonstrate the effectiveness of sensor fusion for estimating the turning velocity and translation motion compensation of the vehicle for generating high-resolution ISAR images. 
\section{Conclusion}
\label{sec:Conclusion}
At short ranges, wideband radars capture multiple detections from a spatially large extended radar target. When these targets undergo rotation within the FoV of the radar, large synthetic apertures are generated for a single radar channel. The wideband long-duration data can subsequently be processed to form ISAR images that provide a top-view or bird's eye view of the vehicle. In conjunction with the front view of camera images, these images can potentially facilitate complete 3D visualizations of an object. However, translation motion compensation must be performed and the Doppler to crossrange axis mapping must be carried out in order to generate meaningful ISAR images. While prior works have focused on blind motion compensation techniques, here, we leverage the availability of vision data in conjunction with radar data for estimating the target state. Specifically, vision data facilitate accurate object detection of the vehicles in the sensor FoV which is useful for eliminating clutter and multipath. Second, they facilitate accurate predictions of the lateral position of the target while radar data enable range estimates. Our work has assumed a constant turn rate model of the target which is typical in road turns. If on the other hand, the target undergoes constant acceleration, then the constant turn rate and acceleration model can be utilized in the extended Kalman framework. Once, the acceleration is computed, then using the centroid tracking method, motion compensation can be carried out by incorporating higher-order components for range alignment. The performance of the sensor fusion for ISAR imaging is evaluated with simulation data generated from a cuboid model of a moving vehicle and from measurements carried out with a millimeter wave radar and a smartphone camera.
The main challenges in the proposed approach lie in handling scenarios where the target moves outside the FoV of either one of the sensors. With suitable upgrades to the sensor tracking model to handle missing data, we believe that complete 3D imaging can be carried out even when the target disappears temporarily from the FoV of any one of the sensors. This will form the focus of our future work. 
\bibliographystyle{ieeetran}
\bibliography{main}
\appendix
\subsection{Performance Analysis of Fusion Algorithm in Complex Environments}
Even when the target is within the FoV of a sensor, the target may not be detected due to environmental conditions (poor light, fog, rain, presence of high clutter) and instrumental defects. In this section, we analyze how the detection performances of the sensors impact the state estimation of the fusion algorithm. We consider the SSUT trajectory of the target vehicle since the vehicle lies within the FoV of both the sensors for almost the entire duration of its trajectory. 

\begin{figure}
\centering
\includegraphics[width = 2.5in, height = 4in]{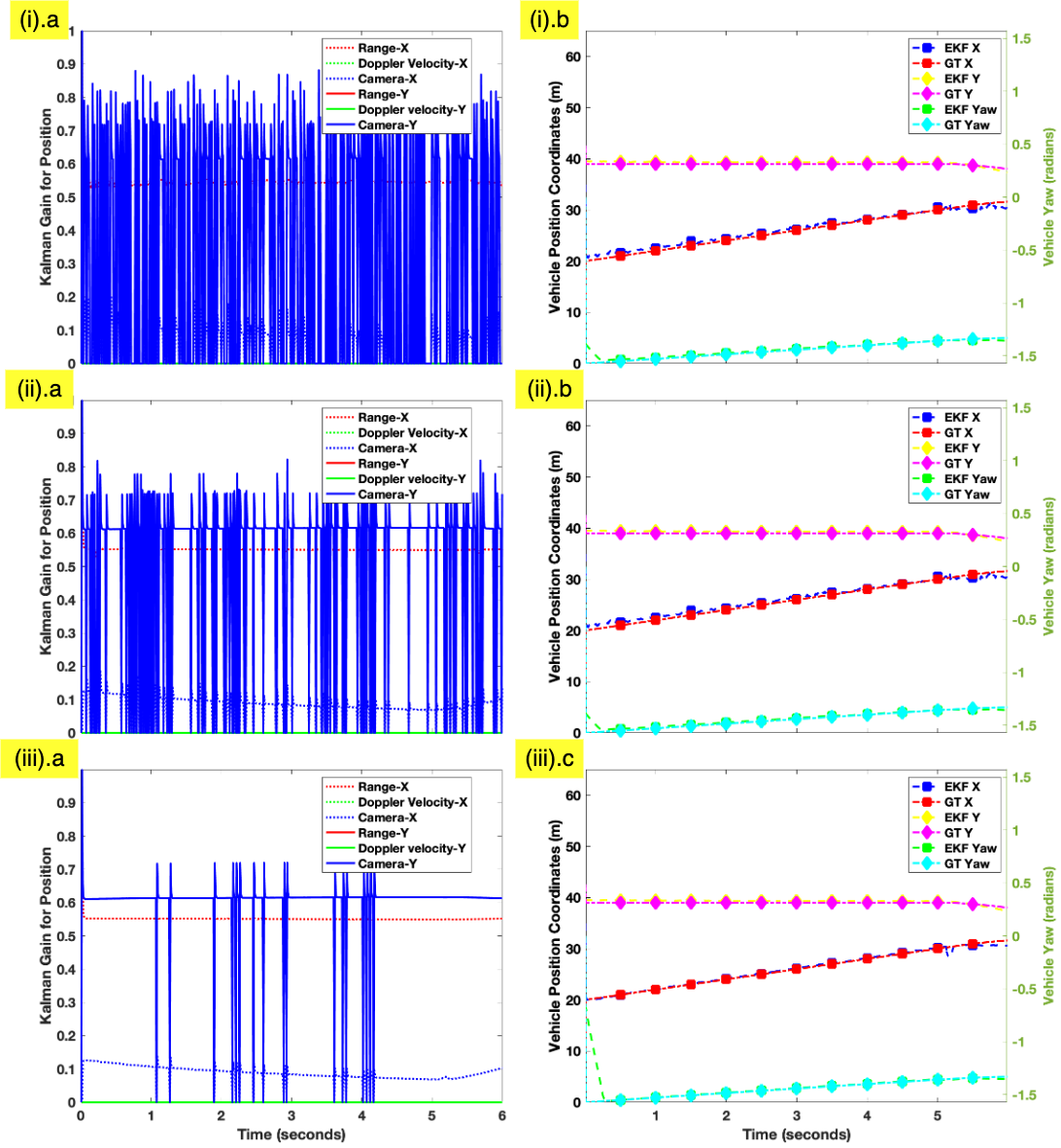}
\caption{Left figures show the Kalman gain for position estimates based on sensor measurements while the right figures show the position and yaw rate state estimation. The top, middle, and bottom rows correspond to results obtained with 50\%, 75\%, and 99\% detection performance of the camera. Radar is assumed to have a high probability of detection (above 90\%) and a low false alarm rate (below $10^{-7}$).}
\label{fig:EKFResults_CamPd}
\end{figure}

\emph{Camera:} First we consider the scenario when the radar is properly functioning throughout the duration of the trajectory while the camera failure is random. In other words, even when the target vehicle is within the camera's FoV, it will be detected only in some of the time frames. The results for Kalman gain and the state estimation for $P_d = 50\%, 75\%$ and $99\%$ are shown in Fig.\ref{fig:EKFResults_CamPd}. 
The top row shows the Kalman gain for position estimates and yaw as a function of the three sensor measurements (range, Doppler and camera-based lateral position of object). 
We observe that $\mathbf{K}_{1,1}$ is higher than $\mathbf{K}_{1,2}$ indicating that along this particular trajectory, the range is more strongly correlated to $\xe$ of the target vehicle than $\ye$. The values never quite fall to zero since the radar measurements are continuously obtained. 
On the other hand, $\mathbf{K}_{2,1}$ and $\mathbf{K}_{2,2}$ are very low (close to zero) throughout the duration of the trajectory indicating that the position of the target is not strongly correlated to the Doppler velocity measurements. $\mathbf{K}_{3,1}$ is also zero throughout the duration of the motion since the camera does not directly provide depth information. However, $\mathbf{K}_{3,2}$ is high because the lateral position estimated by the camera is directly used to estimate $\ye$. When the camera detection fails, $\mathbf{K}_{3,2}$ falls to zero. The frequency with which it falls to zero is greatest in Fig.\ref{fig:EKFResults_CamPd}i.a when $P_d=50\%$ and lower for Figs.\ref{fig:EKFResults_CamPd}i.b and i.c. 
\begin{figure*}[ht]
\centering
\includegraphics[width = 6in, height = 4in]{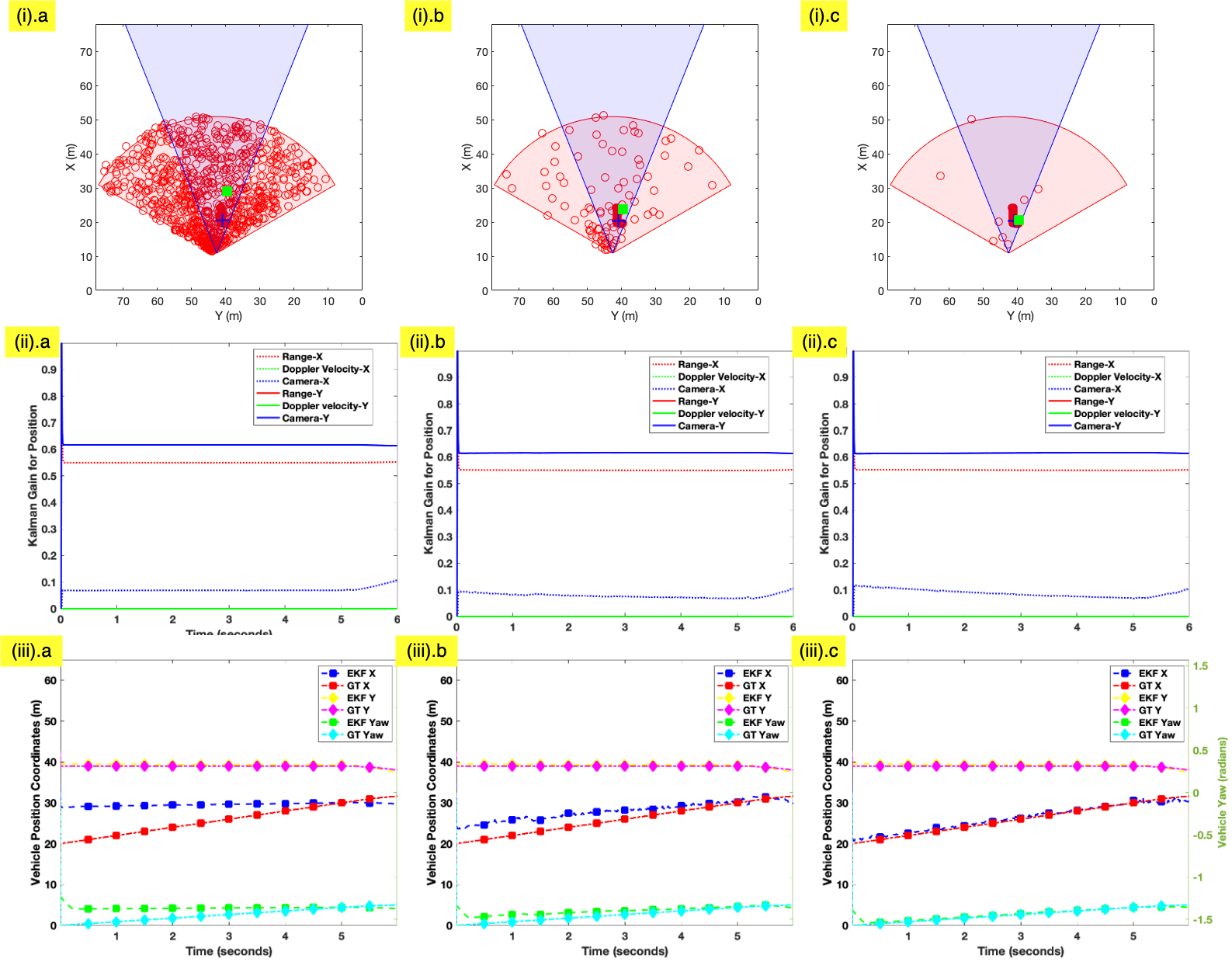}
\caption{Top row shows radar detections (red circles) within the radar coverage area (red region), camera detections (blue plus) within the camera coverage area (blue region), and estimated target centroid (green square). The second row shows the Kalman gain for position estimates based on sensor measurements. The third row shows the position and yaw rate estimation. The columns a, b and c corresponds to radar detection metrics of (a) $P_D = 50\%, P_{fa}=10^{-5}$, (b) $P_D = 75\%, P_{fa}=10^{-6}$ and (c) $P_D = 90\%, P_{fa} = 10^{-7}$.} 
\label{fig:EKFRadRes}
\end{figure*}
The corresponding state estimation results are presented in the second row. The results show that despite the camera's missed detections, the state estimations (positions, yaw) are fairly accurate due to the reliance on predictions from the prior state in the Kalman model. 

\emph{Radar:} We incorporate discrete clutter scatterers in the radar FoV which are assumed to be uniformly spatially distributed while the number of scatterers follows the binomial distribution. Further, the detection of each clutter scatterer in each resolution cell is assumed to be independent across time. Since the radar performs high resolution ranging, we model its performance through the probability of detection ($P_d$) of each point scatterer on the target and the probability of false alarms ($P_{fa}$), which together are a function of the SNR and clutter. The results are presented in Fig.\ref{fig:EKFRadRes}. The top row of the figure shows the visualization of the radar and camera coverage areas in red and blue respectively for a single time frame $0.05s$ for different radar detection metrics. The radar provides multiple detections shown as red circles while the camera provides a single detection shown as a blue plus. The Kalman filtered two-dimensional position estimate of the target vehicle is shown as a green square. 

First, we consider the scenario where $P_d = 50\%$ and the false alarm rate is $P_{fa} = 10^{-5}$ which corresponds to very low SNR resulting in lots of radar detections as seen in Fig.\ref{fig:EKFRadRes}i-a. Th resulting Kalman gain plot in Fig.\ref{fig:EKFRadRes}i-b shows that the radar-based range gain value ($\mathbf{K}_{1,1}$) is fairly stable and not zero. Likewise for camera-based $\mathbf{K}_{3,2}$. The remaining values are low. However, the state estimations in Fig.\ref{fig:EKFRadRes}i-c show high error in $\xe$ due to poor SNR. The $\ye$ is still fairly accurate due to the reliable measurements from the camera. This results in overall deterioration in the yaw estimation. In Fig.\ref{fig:EKFRadRes}ii.a-c, the radar detection metrics correspond to $P_d=75\%$ and $P_{fa}=10^{-6}$. This results in fewer radar detections. The Kalman gain plot in Fig.\ref{fig:EKFRadRes}ii.b remains largely unchanged but there is still a significant error in $\xe$ as seen in Fig.\ref{fig:EKFRadRes}ii.c. When the detection metrics further improve ($P_d=90\%$ and $P_{fa} = 10^{-7}$), then the number of radar detections outside of the target are very few as observed in Fig.\ref{fig:EKFRadRes}iii.a and the state estimation in Fig.\ref{fig:EKFRadRes}iii.c is very accurate. %To summarize, the performance analysis shows that $\xe$ and $\ye$ are strongly correlated to the radar range and camera's lateral position measurements. 
To summarize, $\xe$ is sensitive to the errors due to false alarms in the case of radar while $\ye$ is sensitive to missed detection of the object by the camera. Thus both the sensors are indirectly required for the accurate estimation of the yaw rate which is subsequently used for generating the ISAR images.
\par\noindent
\section*{Acknowledgments}
The research was funded by the MEITY CC\&BT 13(30)/2020 grant from the Govt. of India, Ministry of Electronics and Information Technology. S. S. R. gratefully acknowledges support from Neeraj Pandey, doctoral student at IIIT Delhi for the measurement data collection.
\par\noindent
\begin{IEEEbiography}[{\includegraphics[width=1in,height=1.25in,clip,keepaspectratio]{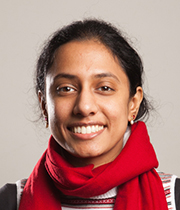}}]%
{Shobha Sundar Ram} is Associate Professor, Dept. of Electronics and Communications Engineering, Indraprastha Institute of Information Technology (IIIT), Delhi. She did her Bachelor of Technology in ECE from the University of Madras, India in 2004 and then her Master of Science and Ph.D. in electrical engineering from the University of Texas at Austin, USA in 2006 and 2009 respectively. Before joining IIIT Delhi, she worked as a research and development electrical engineer at Baker Hughes Inc. USA. She is engaged in research and education principally in the areas of radar signal processing and electromagnetic sensor design and modeling. She is a Senior Member of IEEE, an active member of the Aerospace and Electronics Systems Society and an Associate Editor for the IEEE Transactions on
Aerospace and Electronics Systems.
\end{IEEEbiography}
\end{document}